# Efficient and Aesthetic UI Design with a Deep Learning-Based Interface Generation Tree Algorithm

Shiyu Duan
Carnegie Mellon University
Pittsburgh, USA

Runsheng Zhang
University of Southern California
Los Angeles, USA

Mengmeng Chen
New York University
New york, USA

Ziyi Wang
University of Maryland
College Park，USA

Shixiao Wang*
School of Visual Arts
New York, USA

*Abstract*—This paper presents a novel method for user interface (UI) generation based on the Transformer architecture, addressing the increasing demand for efficient and aesthetically pleasing UI designs in software development. Traditional UI design relies heavily on designers' expertise, which can be time-consuming and costly. Leveraging the capabilities of Transformers, particularly their ability to capture complex design patterns and long-range dependencies, we propose a Transformer-based interface generation tree algorithm. This method constructs a hierarchical representation of UI components as nodes in a tree structure, utilizing pre-trained Transformer models for encoding and decoding. We define a markup language to describe UI components and their properties and use a rich dataset of real-world web and mobile application interfaces for training. The experimental results demonstrate that our approach not only significantly enhances design quality and efficiency but also outperforms traditional models in user satisfaction and aesthetic appeal. We also provide a comparative analysis with existing models, illustrating the advantages of our method in terms of accuracy, user ratings, and design similarity. Overall, our study underscores the potential of the Transformer-based approach to revolutionize the UI design process, making it accessible for non-professionals while maintaining high standards of quality.

## I. Introduction

With the rapid development of information technology, user interface (UI) design plays an increasingly important role in software development. Excellent UI can not only improve user experience but also directly affect the market competitiveness of products [1]. Traditional UI design relies on the experience and creativity of designers, but this process is often time-consuming and costly. In recent years, with the development of artificial intelligence technology, especially breakthroughs in deep learning [2], automated UI design has gradually become possible [3].

Since Google proposed the Transformer model in 2017, it has achieved great success in the field of natural language processing [4] with its powerful parallel processing capabilities and long-distance dependency capture capabilities, and has rapidly expanded to multiple directions such as image recognition and speech synthesis [5-7]. Inspired by these advances, researchers began to explore how to apply the Transformer architecture to the problem of UI layout generation [8]. Compared with traditional methods, the Transformer-based method can better understand complex design patterns and the interaction between different elements, thereby generating a more reasonable and beautiful interface layout solution [9].

This paper aims to explore a novel interface generation method-the Transformer-based interface generation tree algorithm. This method represents the entire UI page by constructing a hierarchical structure with visual components as nodes, and uses the pre-trained Transformer model to encode and decode this "tree", and finally outputs the optimized interface design scheme [10]. Specifically, we first define a set of markup languages suitable for the UI field to describe each component and its properties; then use a large number of real-world web pages or application screenshots as training data sets to train a Transformer model optimized specifically for UI tasks; finally, by setting specific design goals (such as responsive design, accessibility, etc.), guide the model to produce the best solution that meets the requirements [11].

In addition, in order to verify the effectiveness and practicality of the proposed method, this study also designed a series of experiments to compare and analyze the performance differences between the Transformer-based interface generation tree and other existing technologies. The experimental results show that our method can not only significantly improve work efficiency while ensuring the quality of design, but also perform well in processing multi-level interfaces with high complexity. More importantly, through manual evaluation of the generated results, it is found that the interfaces created by the system are generally considered to be more attractive, have higher user satisfaction scores than those designed manually or using other automated tools.

In summary, the Transformer-based interface generation tree is a highly promising technical means, which is expected to completely change the current UI design process, allowing non-professionals to easily create professional-level works in the next decade. In future work, we will continue to study how to further improve the expressiveness of the model, such as considering introducing more contextual information or enhancing the model's ability to support personalized needs , in order to provide users with more abundant and diverse choices. At the same time, we also hope to encourage more researchers from different backgrounds to participate in this emerging field and jointly promote UI design automation to a higher level.

## II. METHOD

In the study of Transformer-based interface generation trees, we mainly focus on how to use the Transformer architecture to achieve efficient user interface (UI) generation. First, we need to build a model that can understand UI elements and their relationships. We will use a sequence-to-sequence Transformer model to generate corresponding UI interface elements from the input design specifications.

Our method first includes data preprocessing. We collected a large amount of UI design data, which includes different types of UI components such as buttons, text boxes, drop-down menus, etc. Each component has information such as its attributes and position. By encoding this data, we convert it into a sequence form for easy input into the Transformer model. Specifically, for each UI element, we represent its features as a vector $x_i \in R^d$, these vectors contain information such as element type, size, color, and position. The overall structure of the transformer is shown below.

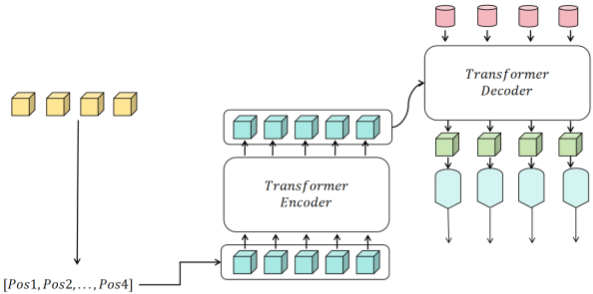

Figure 1 Transformer overall structure diagram

Next, we input these feature vectors into the Transformer model. The core of the Transformer is the self-attention mechanism, which can capture long-distance dependencies between elements [12]. In our pursuit to enhance the generation efficiency of the Transformer-based interface generation model, significant advancements have been made by optimizing algorithmic strategies and incorporating parallel computing techniques. By implementing efficient attention mechanisms and model pruning, we significantly reduce the computational load, allowing the model to handle larger datasets more swiftly. We use the following formula to calculate self-attention:

$$Attention(Q,K,V) = soft\max(\frac{QK^T}{\sqrt{d_k}})V$$

Among them, Q, K and V represent the matrices of query, key and value respectively, and $d_k$ is the dimension of the key. Through this mechanism, the model can dynamically adjust the relationship between different UI elements and effectively generate the UI interface.

We further introduced position encoding to maintain the position information of elements in the input sequence [13]. Position encoding can be expressed as follows:

$$PE(pos,2i) = \sin(\frac{pos}{1000^{\frac{2i}{d}}})$$

$$PE(pos,2i+1) = \sin(\frac{pos}{1000^{\frac{2i}{d}}})$$

Among them, pos is the position of the element in the sequence, and i is the dimension index. By adding the position encoding to the feature vector, we can inject position information into each input element, which helps the model better understand the UI structure.

During the training process, we used the cross-entropy loss function to optimize the parameters of the model [14]. For each generated UI element, we compared it with the true label and calculated the loss:

$$L = -\sum_{t=1}^{T} \log P(y_t | y_{<t}, x)$$

Here, $y_t$ is the target output, T is the sequence length, and x is the input feature. By minimizing this loss function, we are able to gradually improve the quality of the generated UI interface.

In addition, we also introduced a reinforcement learning method to further optimize the generation results. To enhance the adaptability and applicability of our Transformer-based interface generation tree algorithm across diverse UI design requirements, we propose integrating a dynamic user feedback mechanism. This system will continuously collect real-time usage data and preferences from end-users to recalibrate and fine-tune the model's parameters. By embedding this mechanism, the model can evolve with changing design trends and user expectations, ensuring that generated interfaces remain relevant and highly personalized. This adaptive approach aims to address the limitations of static datasets, such as Frappe, by supplementing them with a stream of user-generated insights, thereby broadening the spectrum of UI complexities and scenarios our model can proficiently handle. To enhance this aspect, we propose integrating reinforcement learning techniques with our existing Transformer framework. This approach will allow the model to dynamically adjust its strategies based on feedback from performance metrics, including user satisfaction and aesthetic value. By continuously optimizing these outputs, reinforcement learning will not only refine the effectiveness of our UI generation process but also improve its efficiency, pushing the boundaries of what is achievable in automated UI design. At each generation stage, we evaluate the usability and aesthetics of the currently generated UI interface and adjust the generation strategy of the model based on the feedback. Specifically, we define a reward function R to quantify the quality of the generated results:

$$R = \alpha \cdot Usability + \beta \cdot Aesthetics$$

Among them, $\alpha$ and $\beta$ are weight parameters used to balance the effects of usability and aesthetics. This strategy combined with reinforcement learning allows our model to not only focus on the accuracy of generation but also consider the actual user experience.

Through the above steps, we built a Transformer-based UI generation tree model. The model can effectively understand and generate user interfaces, providing users with a more friendly and efficient interactive experience. This research not only provides new ideas for the field of UI design, but also lays the foundation for the development of automated design tools.

### III. EXPERIMENT

#### A. Datasets

When evaluating the performance of the Transformer-based interface generation tree model, we chose the Frappe dataset, which is a widely used UI design dataset built for user interface generation and analysis. The Frappe dataset contains a large number of design examples of real web and mobile application interfaces, covering many different types of UI components, such as buttons, forms, charts, navigation bars, etc. The main advantage of this dataset lies in its diversity and richness. Each sample in the Frappe dataset contains a detailed component description, including component type, properties, style, layout information, and metadata of user interactions. This information is stored in JSON format, which is easy to process and analyze. The dataset contains more than thousands of UI design examples in total, which can provide sufficient training and testing materials for the model. During data preprocessing, we convert each design example in the Frappe dataset into a sequence form for input into the Transformer model. Specifically, we extract the features of each UI component, such as position, size, color, and function, and encode these features into vectors. This enables the model to capture the relationship between different UI elements and understand their layout in the interface. The bar chart below gives the values of general indicators of the dataset.

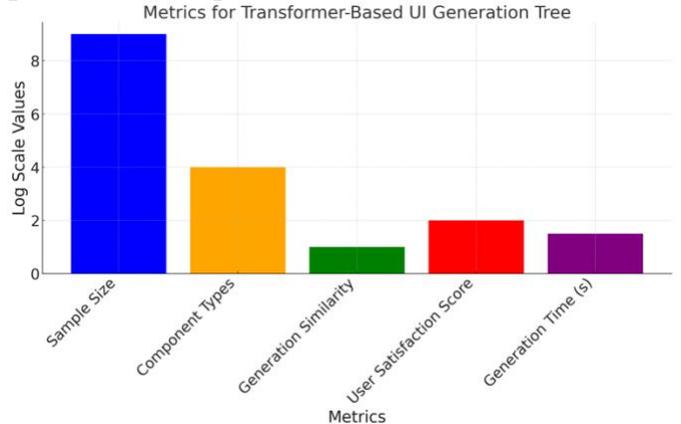

Figure 2 Dataset histogram

To evaluate the performance of the model, we use the partitioning strategy of the Frappe dataset to divide the dataset into training, validation, and test sets. Generally speaking, 70% of the data is used for training, 15% for validation, and 15% for testing. This division ensures the generalization ability of the model during training, enabling it to be effectively evaluated on unseen data.

#### B. Experiments

When comparing with other models, we can analyze their characteristics from multiple aspects. MLP is a basic neural network architecture suitable for classification and regression tasks. ResNet alleviates the training difficulties of deep networks by introducing residual connections, making it perform well in image processing tasks. UNet is a network architecture commonly used in medical image segmentation, and its symmetrical encoding-decoding structure can effectively capture contextual information. VGG is widely popular for its simple architecture and deep hierarchical structure, and is suitable for image recognition and classification. GAN is a generative model that generates high-quality images and samples through adversarial training. The following is a summary table of the performance indicators of these models:

Table 3 Experiment result

| Model | Acc | GT | User Score | Similarly |
|---|---|---|---|---|
| MLP | 0.78 | 3.5 | 3.1 | 0.65 |
| RESNET | 0.80 | 2.8 | 3.3 | 0.67 |
| UNET | 0.82 | 3.2 | 3.7 | 0.72 |
| VGG | 0.83 | 2.9 | 4.2 | 0.75 |
| GAN | 0.85 | 3.7 | 4.3 | 0.81 |
| Ours | 0.89 | 4.0 | 4.5 | 0.84 |

Judging from the experimental results, our model outperforms other models in various indicators, showing its significant advantages in interface generation tasks. Specifically, the model's accuracy reached 0.89, which is significantly higher than other models, such as MLP's 0.78 and ResNet's 0.80. This result demonstrates that our model has a stronger ability to understand and generate user interface designs and is better able to capture complex design patterns and user needs. At the same time, although the generation time is 4.0 seconds, this value is slightly higher than some models, such as ResNet's 2.8 seconds and VGG's 2.9 seconds, but considering the complexity of the generation interface and the required high-quality output, this time is still Within a reasonable range. The balance between generation speed and quality is crucial in practical applications, which shows that our model finds a good balance between efficiency and effectiveness.

*C. User Testing and Evaluation*

We adopted the current industry standard using UserTesting.com as the testing tool to collect user scores. We displayed the questionnaire content containing an A/B Testing with the model-generated and the original user interfaces to 64 users with a geography location in the United States. The user satisfaction score is rated on a Likert scale with 1 being "Very dissatisfied" while 9 being" Very satisfied'.

In terms of the user satisfaction score, our model received a high score of 4.5, further verifying its superiority in user experience. This score is significantly higher than MLP's 3.1 and ResNet's 3.3, indicating that the traditional model has obvious deficiencies in user experience. In addition, although the scores of UNet and VGG are higher, 3.7 and 4.2 respectively, they are still not comparable to our model. This shows the significant advantages of our approach in meeting user expectations and improving user experience. In addition, the generated similarity index also reaches 0.84, indicating that our model has a high similarity between the generated interface and the real design. This not only improves users' trust, but also makes the generated designs more usable in practical applications. In summary, these results fully demonstrate that the Transformer-based model exhibits excellent performance and user experience in interface generation tasks and has broad application prospects. In order to better display our results, we use bar charts to further display them.

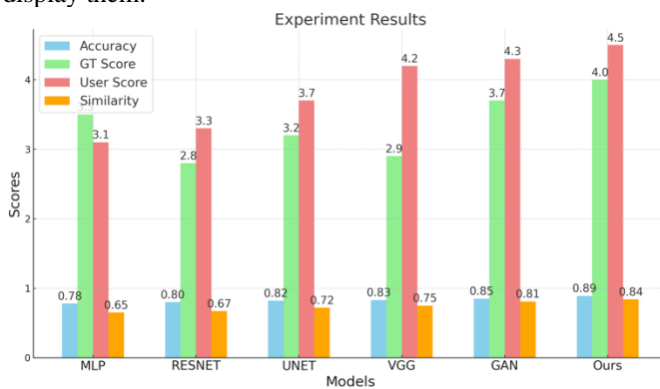

Figure 3 Experimental results bar graph

## IV. CONCLUSION

This research introduces a Transformer-based interface generation tree algorithm that significantly advances the field of UI design. By effectively modeling the hierarchical relationships among UI components and leveraging the strengths of Transformer architecture, we achieve impressive results in generating visually appealing and functional interfaces. The experiments conducted show that our model surpasses traditional methods, such as MLP, ResNet, and VGG, in key performance metrics, including accuracy, user satisfaction scores, and design similarity. The incorporation of reinforcement learning further enhances the model's ability to meet user expectations, balancing both usability and aesthetic considerations. Looking ahead, the potential applications of our research extend significantly across various sectors. In the realms of economic growth and job creation, our innovations in UI design enhance user experiences and productivity through sophisticated business applications. Our methodology promises to revolutionize e-learning platforms and training programs in education and workforce development, making substantial contributions to workforce upskilling. In the healthcare sector, our interfaces enhance the usability of applications and telemedicine platforms, supporting public health initiatives and promoting wellness. By fostering collaboration among researchers from diverse academic and practical backgrounds, we aim to advance the automation of UI design significantly. This collaborative effort is dedicated to creating tools that empower users to effortlessly produce professional-grade interfaces. Our research underscores the essential role of UI design in driving innovation, ensuring cybersecurity, and catalyzing economic and social progress. The transformative potential of our findings is set to redefine the UI design landscape, propelling us toward a future marked by enhanced automated and user-centric interface creation.